# An Improved Physically-Based Surface Triangulation Method

Shangyu Lei[1,†], Wei Fan[1,‡], Hui Ren[1,*]

[1]Institute of Aerospace Vehicle Dynamics and Control, School of Astronautics,

Harbin Institute of Technology, Harbin, 150001, China

**Abstract**

This paper proposes improvements to the physically-based surface triangulation method, bubble meshing. The method simulates physical bubbles to automatically generate mesh vertices, resulting in high-quality Delaunay triangles. Despite its flexibility in local mesh size control and the advantage of local re-meshing, bubble meshing is constrained by high computational costs and slow convergence on complex surfaces. The proposed approach employs conformal mapping to simplify surface bubble packing by flattening the surface onto a plane. Surface triangulation is induced from the planar mesh, avoiding direct bubble movement on the surface. Optimizing bubble quantity control and separating it from the relaxation process accelerates convergence, cutting computation time by over 70%. The enhanced method enables efficient triangulation of disk topology surfaces, supports local size control, curvature adaptation, and re-meshing of discrete surfaces.

## 1  Introduction

The solution of partial differential equations (PDEs) on surfaces or solids with complex topology and geometry typically involves the use of methods including the finite element method (FEM), finite volume method, and isogeometric analysis. These methods rely on mesh generation. In FEM, the common use of linear finite elements requires surface triangulation, which is also critical to volumetric triangulation. The surface triangulation algorithm requires maximizing the minimum angle, as mesh quality is crucial for the accuracy and convergence of the finite element solution. Additionally, the mesh vertex density should be adjustable to local geometric or physical fields. Common mesh generation methods include Delaunay triangulation, advancing front techniques, quadtree-based approaches, and Voronoi polygon-Delaunay methods. Many studies also focus on improving mesh quality to enhance the accuracy of solutions.

### 1.1  Bubble Meshing

Bubble meshing is a physically-based method for obtaining the globally optimal positions of mesh vertices. In this approach, In this approach, bubbles are treated as freely moving entities within a geometric domain, where attractive and repulsive forces between them result in an even distribution across the domain. The center of each bubble represents a mesh vertex, and the bubble's radius represents the distance between mesh vertices. In the 1990s, Shimada introduced the bubble meshing method[1]. Shimada and Gossard[2] later proposed a bubble meshing technique for trimmed parametric surfaces. Subsequently, Shimada and his colleagues developed algorithms for generating anisotropic triangular[3] and tetrahedral[4] meshes by filling ellipsoids.

[†]Email: 24s018022@stu.hit.edu.cn

[‡]Email: fanwei@hit.edu.cn

[*]Email: renhui@hit.edu.cn

The excessive iterations in the dynamic adjustment process of bubble meshing lead to a reduction in computational efficiency. Ying Liu et al.[5] demonstrated the inherent parallelism of bubble meshing through computational experiments. Research by Qi et al.[6] indicated that more than 80% of the total simulation time is consumed by the calculation of new bubble positions and bubble quantity control (dynamic adjustment). This finding highlighted key areas for improving computational efficiency and provided several acceleration strategies, such as optimizing the initial bubble density and allowing the viscosity coefficient to change over time and space to speed up convergence. Guo et al.[7] later proposed additional acceleration strategies. Li et al.[8] studied the conditions for generating high-quality meshes using the bubble meshing method and their superconvergence properties. Most of these strategies focus on optimizing the original method, while the basic strategy of dynamic adjustment remains unchanged, continuing to alternate between bubble quantity control and physically-based relaxation.

Surface triangulation with bubble meshing is achieved through two methods. The first method directly generates the mesh in 3D space, allowing bubbles to move on the surface[1, 2]. During node relaxation on the surface, surface attraction can be added to prevent nodes from detaching from the surface[9, 10]. While suitable for various complex surfaces, it relies on surface parametric equations to perform complex calculations, such as partial derivatives[1, 2], geodesic distances[11], and determining surface attraction on nodes[9, 10].

The second method generates the mesh in the parametric domain and then maps it onto the surface. Zheleznyakova et al.[12, 13] treated mesh vertices as charged particles and used molecular dynamics simulations to move them to optimal positions in the parametric domain of a NURBS surface. However, this mapping approach may introduce angular distortion when the parametric mesh is mapped back onto the surface. To address this, Wang et al.[14] introduced three local adjustments to the mesh generation algorithm to mitigate the distortion.

**1.2 Conformal Parameterization**

Mesh parameterization maps a surface embedded in three-dimensional space to a simpler parametric domain, enabling tasks such as texture mapping, mesh deformation, editing, and re-meshing. To avoid angular distortion, conformal mapping establishes an angle-preserving correspondence between the surface and the parametric domain. Conformal parameterization is divided into two categories: parameterization for topological disks and global parameterization for high-genus surfaces. The first category includes several approaches: Cauchy-Riemann equation approximation: Levy et al.[15] computed a piecewise linear approximation of the Cauchy-Riemann equations using a least-squares method; Harmonic energy minimization: Desbrun et al.[16] computed conformal parameterization by minimizing the Dirichlet energy of the surface; Angle-based methods: Sheffer et al.[17] introduced an angle-based parameterization method that directly optimizes the angles of the flattened mesh to approximate the angles of the input surface mesh. Zayer et al.[18] linearized this process; Laplace operator linearization: Haker et al.[19] used the discrete Laplace operator to convert a PDE into a linear system to compute global conformal mappings from genus-zero surfaces to a sphere. Sawhney and Crane[20] proposed a linear conformal parameterization method that can conformally flatten a surface onto any given target shape. The methods mentioned above for topological disks require surface cutting when dealing with high-genus surfaces. The second category allows for global parameterization of high-genus surfaces without cutting[21]. The angle-preserving property and the simplicity of the parametric domain make conformal parameterization an effective method for surface mesh re-meshing. Re-meshing improves the input mesh by adding, deleting, or moving mesh vertices to meet specific quality requirements. After the surface is conformally flattened, re-meshing can be performed in the planar domain[22, 23]. The advantage of the sphere packing method in generating high-quality meshes makes it an effective approach for re-meshing[24, 25].

Bubble meshing generates high-quality meshes with flexible control over local mesh size, shape, and re-meshing. However, its surface method requires many iterations and complex calculations, limiting its computational efficiency and applicability. This paper aims to simplify the method and improve its computational efficiency while preserving the advantages of the original approach. Chapter 2 introduces the basic method of surface triangulation using conformal mapping. Chapter 3 presents a new bubble quantity control method to accelerate mesh generation in the parametric domain. Chapter 4 discusses conformal mapping and re-meshing techniques, with a focus on adjusting bubble size in re-meshing. Chapter 5 provides a curvature-based method for controlling mesh size.

## 2 Algorithm Theory

### 2.1 Conformal Parameterization

A conformal mapping is an angle-preserving mapping. A mapping $f$ from a two-dimensional manifold $M$ to the complex plane $\mathbb{C}$ is called conformal if it preserves the angles between two vectors, while allowing changes in the magnitude of the vectors. One key property is that it transforms infinitesimal circles into infinitesimal circles while maintaining angles, known as the circularity-preserving property.

### 2.2 Surface Bubble Mesh Method Based on Conformal Mapping

Smooth surfaces can be subdivided into geodesic triangles, where each edge is a segment of a geodesic line on the surface. By replacing the geodesic edges with straight lines and the geodesic triangles with Euclidean triangles, a polyhedral discrete surface is created, approximating the original smooth surface. The surface bubble meshing method generates high-quality geodesic triangulation by tightly packing spheres on the surface. Conformal mapping projects the surface onto the complex plane, with the preservation of both the geodesic Delaunay triangulation and the tangential arrangement of bubbles. This ensures the correspondence between the geodesic triangulation on the surface and the Euclidean triangulation on the complex plane obtained through the mapping. As a result, a surface triangulation method based on conformal mapping is obtained:

Given a smooth surface $M$ in Euclidean space, a conformal mapping $f$ projects it onto the complex plane. Using the bubble meshing method, a triangular mesh $T = (V, E)$ is generated on the plane, where $V$ and $E$ denote the sets of vertices and edges, respectively. The inverse mapping $f^{-1}(T)$ induces a geodesic triangulation on the smooth surface $M$, where $f^{-1}(V)$ represents the sample points on $M$ and $f^{-1}(E)$ represents the geodesic edges on $M$. Replacing the geodesic triangles with Euclidean triangles results in a discrete surface, which serves as a triangulation of the smooth surface $M$, with $f^{-1}(V)$ being the mesh vertices.

Through conformal mapping, the bubble-packing problem on a surface with complex geometry is reduced to a planar bubble-packing problem, allowing for local control over mesh size and shape and re-meshing in the plane. This significantly simplifies the bubble-packing process on the surface. Finally, the inverse mapping of the planar mesh onto the surface yields the final surface triangulation.

The algorithm framework is as follows:

a)   Generate an initial discrete surface by meshing the parametric domain.

b)   Apply a conformal mapping of the initial discrete surface onto the complex plane.

c)   Re-mesh the planar domain to improve mesh quality.

d)   Inversely map the new mesh back to the surface, obtaining the triangulation of the smooth surface.

In contrast to the previous method of directly applying conformal parameterization to the smooth surface, this approach first discretizes the surface, as the conformal mapping algorithm used here (BFF) only supports conformal flattening for discrete surfaces.

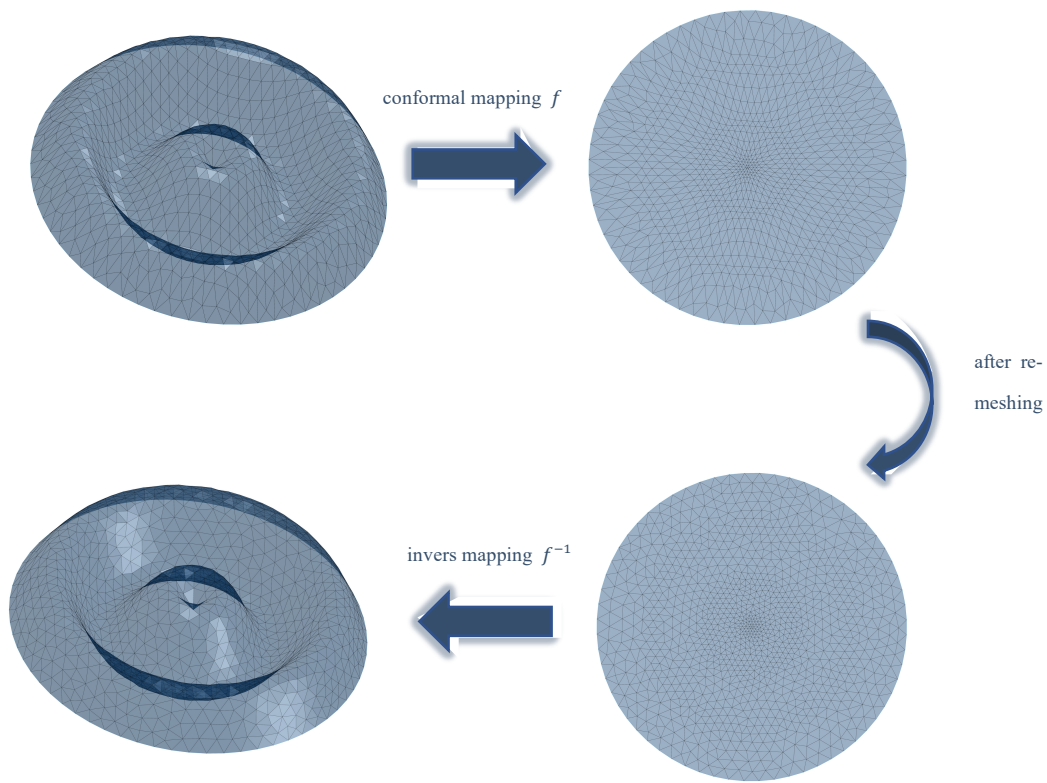

Fig. 1. Process of surface bubble meshing

## 3    Initial Discrete Surface

The basic steps for bubble packing in the planar parametric domain are similar with original planar methods. Bubbles (2D circles) are packed along the boundary and interior of the geometric domain using binary and diagonal quadtree methods[2]. The radii of interior bubbles are interpolated based on both boundary bubbles and pre-inserted interior bubbles[26] (hereafter collectively referred to as anchor bubbles), with closer bubbles assigned higher weights, as shown in Fig. 2. Once packing is complete, the bubble centers are directly connected to form Delaunay triangles, with no need for dynamic adjustments between bubbles. This process is conducted as part of the re-meshing procedure.

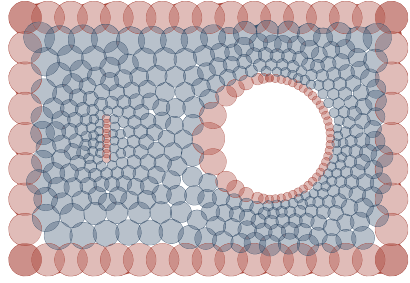

Fig. 2. The radius of interior bubbles (blue) is influenced by anchor bubbles (red).

After triangulating the planar parametric domain, the mesh is mapped onto the surface to obtain the initial discrete surface. The initial discrete surface must be generated with acceptable error during the first bubble packing in the parametric domain. According to the error approximation theorem, let $M \subset \mathbb{R}^3$ be a $C^2$-class smooth surface and T a discrete triangular mesh approximating $M$. There exists a constant $c > 0$ such that the geometric error between T and $M$ satisfies:

$$d_H(T, M) \leq ch^2$$

where $d_H(T, M)$ is the Hausdorff distance between T and $M$, and $h$ is the maximum edge length in T. During the discretization of a smooth surface, controlling the maximum edge length is critical. Regions of higher curvature require denser sampling, meaning smaller bubble radii.

The maximum edge length can be determined by the given allowable error $\varepsilon$ as follows[27]:

$$\frac{s - l}{l} \leq \varepsilon$$

where $s$ is the arc length of a geodesic edge on the surface, and $l$ is the corresponding straight edge length in the Euclidean triangle, as shown in Fig. 3.

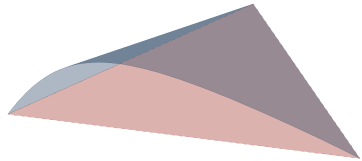

Fig. 3. Geodesic triangle and Euclidean triangle

For any point $P$ on the surface $S$, let $v \in T_p S$ be an arbitrary tangent vector at $P$. The maximum allowable mesh edge length at $P$ in the direction of $v$, denoted $l_v$, is given by[28]:

$$l_v = (1-\varepsilon)\frac{1}{\kappa_n}\sqrt{40(1-(1-1.2\varepsilon)^{0.5})} = \frac{g(\varepsilon)}{\kappa_n}$$

where $\kappa_n$ is the normal curvature at $P$ along the direction of $v$. The maximum allowable edge length at point $P$, denoted $P$, is:

$$l_p = \min\{l_v | v \in T_p S\}$$

In this case, $\kappa_n$ takes its maximum value, and the vector $v$ is aligned with the direction of maximum curvature at point $P$.

Since the maximum allowable edge length is required for bubble packing in the parametric plane, after determining its value $l_p$ in $E^3$-space, it needs to be calculated in the parametric domain. Let $f$ be the continuous mapping from the parametric domain to $E^3$-space, with $r = f(u,v)$, and the Jacobian matrix of $f$ given by $J_f = (f_u, f_v)$. The singular value decomposition (SVD) of the Jacobian matrix is applied:

$$J_f = U \sum V^T = U \begin{pmatrix} \sigma_1 & 0 \\ 0 & \sigma_2 \\ 0 & 0 \end{pmatrix} V^T$$

Where $\sigma_1 \geq \sigma_2 \geq 0$, the mapping $f$ transforms the unit disk in the parametric domain into an ellipse in $E^3$-space, with semi-axes $\sigma_1$ and $\sigma_2$. Thus, the maximum allowable edge length at point $P$ in the parametric domain is:

$$l'_p = l_p / \sigma_1$$

The bubble radius control function in the parametric domain is:

$$r \leq \frac{l'_p}{2}$$

The bubble radius is influenced not only by the anchor bubbles but also by the maximum allowable edge length of the mesh. By applying the bubble radius control function, the radius is restricted in areas with high curvature, enabling denser sampling to capture the surface geometry more accurately. Combined with re-meshing, this approach improves mesh quality and facilitates curvature-adaptive surface triangulation.

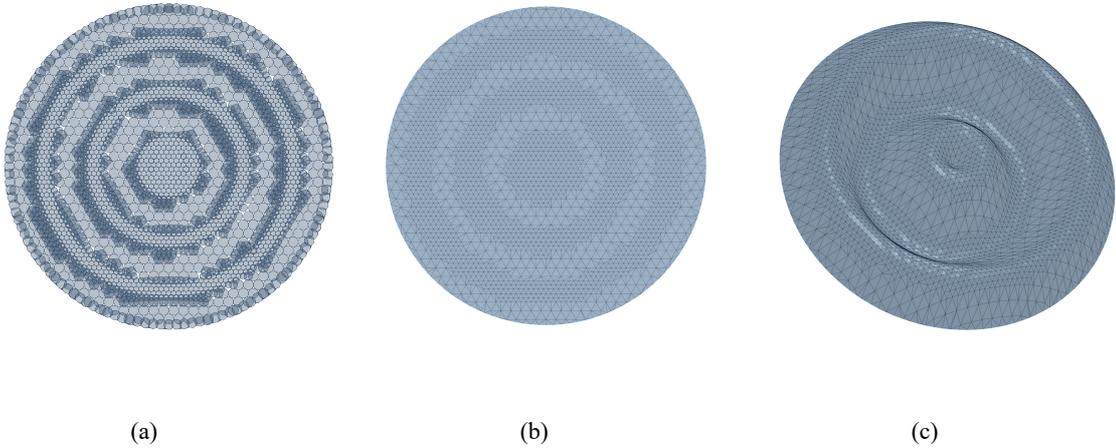

(a)          (b)          (c)

Fig. 4. The triangulation of the parametric domain and the initial discrete surface are controlled by the bubble radius function.

## 4  Surface Re-meshing

The initial discrete surface is obtained by mapping the mesh from the planar parametric domain onto the surface. Since this mapping is not angle-preserving, the surface requires re-meshing to improve the mesh quality.

### 4.1  Conformal Parameterization

This paper uses the boundary-first flattening (BFF) method[20] to achieve conformal parameterization of the discrete surface. The BFF method can conformally flatten domains with disk topology onto any given target shape. Since the re-meshing process imposes no specific requirements on the boundary shape of the flattened mesh, automatic parameterization is applied to minimize area distortion, with the conformal factor set to $\lambda = 0$ at the boundary vertices. The flattened mesh retains the same internal angles as the original discrete surface, making it possible to re-mesh the plane and improve the mesh quality.

### 4.2  Re-meshing

The core of generating high-quality meshes with the bubble-packing method lies in its physically-based relaxation process. Luo[25] proposed a similar approach for optimizing 2D region meshes, where bubbles are generated at mesh vertices and vertex positions are optimized based on inter-bubble forces. The re-meshing method in this paper is inspired by this approach.

The first step is to generate bubbles of appropriate sizes at the mesh vertices, ensuring that the high-quality mesh is preserved during the physically-based relaxation process. The bubble sizes must match the mesh size distribution of the initial discrete surface, allowing the re-meshing process to improve mesh quality without altering the overall mesh size. The radius of each newly generated boundary bubble is:

$$2r_j = \frac{1}{2}(l_{ij} + l_{jk})$$

where $i, j, k$ are consecutive vertices on the boundary. These boundary bubbles are tangential to each other, ensuring that bubbles do not cross the boundary during the physical relaxation process. The diameter of an interior bubble at vertex iii is calculated as the weighted average of the lengths of edges $l_{ij}$ connected to vertex $i$, where $j$ represents the neighboring bubble of $i$. The weight $\omega_j$ is given by the normalized inverse of the edge lengths:

$$\begin{cases} r_i = \frac{1}{2} \sum_{ij \in E} \omega_j l_{ij} \\ \omega_j = \frac{1/l_{ij}}{\sum_{ik \in E} 1/l_{ik}} \end{cases}$$

Using the inverse of edge length as the weight reduces the impact of longer edges on bubble size. The bubble radius conforms to shorter edges and avoids becoming too large, which could erase short edges and result in the loss of surface geometry.

Following the reconstruction of bubbles at the vertices, additional bubbles are placed in the gaps caused by mesh stretching. The radii of these bubbles are interpolated from the reconstructed boundary and interior bubbles. When conformal mapping transforms infinitesimal circles into infinitesimal circles, area scaling occurs, measured by the conformal factor $\lambda$. The relations are $ds' = e^\lambda ds$ and $dA' = e^{2\lambda} dA$, and in the discrete case, $\lambda_{ij} = l'_{ij}/l_{ij}$ . The radii of bubbles, based on edge lengths, implicitly contain information about area scaling at the vertices, providing a reference for nearby bubble placement.

After filling bubbles within the domain, quantity control and physical relaxation are applied to achieve a uniform bubble distribution. The interaction forces between bubbles are analogous to Van der Waals forces, where the interaction force is zero when two bubbles are tangential. The motion of the bubbles is governed by a control equation based on a mass-spring-damper system[2]:

$$m_i \frac{d^2 \boldsymbol{x}_i(t)}{dt^2} + c_i \frac{d\boldsymbol{x}_i(t)}{dt} = \boldsymbol{f}_i(t) \tag{1}$$

$m_i$ is the mass of the $i$-th bubble, $c_i$ is the damping coefficient, $\boldsymbol{x}_i$ is the position vector of the $i$-th bubble, and $\boldsymbol{f}_i$ is the net force exerted on the $i$-th bubble by other bubbles. The control equation (1) is solved using the fourth-order Runge-Kutta method, adjusting the position of each bubble sequentially. One iteration is completed after the adjustment of all bubbles. This iteration is repeated until the net force on each bubble falls below a specified threshold, a process referred to as physically-based relaxation. In the original method, after several iterations of physical relaxation, bubble quantity control is applied. Excess bubbles surrounding each bubble are removed, and additional bubbles are inserted in gaps to prevent bubbles slowly drifting from dense to sparse regions. The criteria for adding and removing bubbles are as follows[2]:

$$overlap = \sum_{i=1}^{n} \frac{2r_0 + r_i - l_i}{r_0}$$

The overlap of a bubble with radius $r_0$ is determined by the radii $r_i$ of $n$ neighboring bubbles and the distance $l_i$ between their centers. Neighboring bubbles are defined as bubbles with a center distance of no more than $2r_0$. If the total overlap is too small (approximately 500% or less), indicating too few neighboring bubbles, another bubble is inserted to fill the gap. If the total overlap is too large (approximately 800% or more), the bubble is deleted. This process of quantity control alternates with physical relaxation until convergence is achieved. The original method aimed to optimize the mesh topology, ensuring that the number of connected edges for each vertex remained within a certain range. However, for non-uniform meshes or meshes with specific boundary shapes, it was difficult to maintain a vertex degree of 6, as in Fig. 5. In such cases, the conditions for adding or removing bubbles had to be relaxed; otherwise, the initial tangential arrangement of bubbles would be disrupted, preventing convergence.

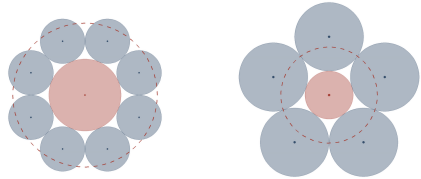

Fig. 5. The $r_0$ bubble is arranged tangentially with neighboring bubbles, but additional bubbles may need to be added or removed around it based on the overlap ratio.

To address this issue, this paper proposes a bubble quantity control strategy specifically for the boundary region. We first introduce the algorithm using planar bubble packing as an example and then extend it to the re-meshing process. When employing the diagonal quadtree method for bubble packing within the domain, the planar domain is subdivided into several rhombuses, with tangential bubbles positioned at the four vertices of each rhombus. This arrangement ensures uniform packing of the initial bubbles, providing a vertex degree of 6 without the need for further adjustment. However, since the boundary shape cannot perfectly align with the discrete edges formed by the rhombuses, excessive overlap between interior and boundary bubbles frequently occurs. Based on this observation, we propose a quantity control method that regulates bubble count according to the degree of overlap near the boundary. For each anchor bubble, the algorithm identifies neighboring bubbles and removes those with excessive overlap. Once all anchor bubbles are processed, the algorithm applies physical relaxation directly, eliminating the need for repeated quantity control. The formula for determining bubble overlap is as follows:

$$overlap = \frac{r_0 + r_i - l_i}{\min\{r_0, r_i\}}$$

After bubble removal, gaps appear near the boundary. To limit the size of gaps, bubbles are typically removed when the overlap exceeds 1, while minimal overlap does not require removal. Following the quantity control in the boundary region, gaps and slight overlaps are present near the boundary bubbles. Initial bubbles packed within the domain using the diagonal quadtree method may also have slight overlaps. During the process of physical relaxation, overlapping bubbles move toward the gaps, and once equilibrium is reached, all bubbles are evenly distributed. The entire process is shown in Fig. 6.

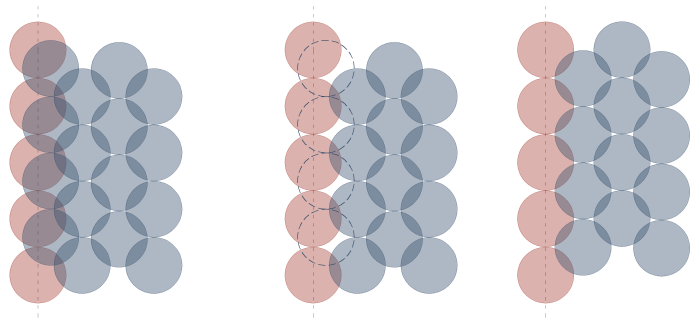

Fig. 6. Bubbles excessively overlapping boundary bubbles are removed, leading to a uniform distribution within the domain after physical relaxation.

The new quantity control method only requires checking neighboring bubbles for anchor bubbles, eliminating the need to repeatedly perform quantity control on all bubbles during the physical relaxation process. This reduces computational cost, decreases the number of iterations in the relaxation process, and speeds up convergence.

After bubble placement, boundary region quantity control and physical relaxation are applied. Upon convergence, bubble centers are connected to form Delaunay triangles. During boundary region quantity control, all bubbles reconstructed at boundary and interior vertices serve as anchor bubbles. Unlike the anchor bubbles on the boundary, which remain fixed during physical relaxation, the interior anchor bubbles move under net forces to achieve a globally optimal arrangement of vertex positions within the domain. The remeshing process is shown in Fig. 7.

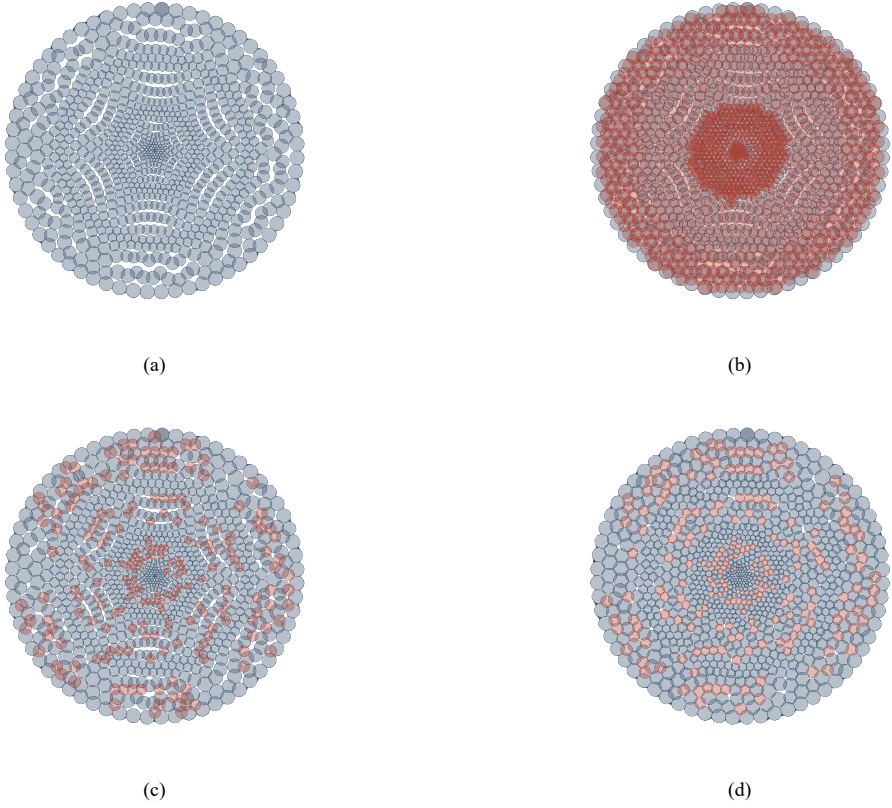

(a) (b)

(c) (d)

Fig. 7. Fig. (a)-(d) illustrate the entire re-meshing process: bubbles reconstructed on the flattened mesh, bubbles inserted in gaps using the boundary region quantity control method, and a uniform distribution achieved after physical relaxation.

## 5    Inverse Mapping

After re-meshing in the planar domain, a piecewise linear mapping is established through barycentric coordinates to map the new mesh back to the original discrete surface. A triangle in the initial discrete surface is denoted as $\Delta = [r_1, r_2, r_3]$, where $r_1, r_2, r_3$ are coordinates defined in 3D Euclidean space. After conformal mapping, its flattened counterpart is denoted as $\widetilde{\Delta} = [u_1, u_2, u_3]$, where $u_1, u_2, u_3$ are coordinates defined in the complex plane. For a point $v_k$ inside the triangle $\widetilde{\Delta}$, with coordinates $\boldsymbol{u} = (u, v)$ in the complex plane, there exist unique scalars $\lambda_1(\boldsymbol{u}), \lambda_2(\boldsymbol{u}), \lambda_3(\boldsymbol{u})$ such that:

$$\boldsymbol{u} = \sum_{i=1}^{3} \lambda_i(\boldsymbol{u})\, \boldsymbol{u}_i \qquad (2)$$

The scalars $\lambda_i(\boldsymbol{u})$ are the barycentric coordinates of the point $v_k$ relative to the triangle $\widetilde{\Delta}$, and $\sum_{i=1}^{3} \lambda_i(\boldsymbol{u}) = 1$. Since conformal mapping preserves angles, the triangles $\widetilde{\Delta}$ and $\Delta$ are similar. The point $v_k$ on $\widetilde{\Delta}$ retains the same barycentric coordinates as its corresponding point on $\Delta$ after the mapping. By replacing $\boldsymbol{u}_i$ with $\boldsymbol{r}_i$ in equation

(2), we obtain the piecewise linear mapping $\Psi: \widetilde{M} \to M$. $\Psi(v_k)$ maps the vertex $v_k$ in the new mesh to the initial discrete surface, and is expressed using barycentric coordinates as:

$$\Psi(v_k) = \sum_{i=1}^{3} \lambda_i(\boldsymbol{u})\, \boldsymbol{r}_i$$

Through piecewise linear mapping, the coordinates of the new mesh vertices on the discrete surface are obtained, providing an approximate inverse mapping to the vertices on the smooth surface. Curvature-based size control ensures that the error between the initial discrete surface and the smooth surface remains within an acceptable range. Consequently, the error between the new mesh and the smooth surface is also controlled, ensuring the accuracy of the approximate inverse mapping.

## 6 Experimental Results

We demonstrate the mesh quality and time consumption achieved with the improved boundary region quantity control method through planar and surface examples, comparing it to the original quantity control method. The program is implemented in C++.

*Rectangular Plate with a Hole:*

Fig. 8 shows the uniformly sized bubbles and mesh generated on a flat plate using the boundary region quantity control method. After filling initial bubbles in the parametric domain, dynamic adjustments were made using both quantity control methods. The final results, shown in Fig. 9, indicate that the new quantity control method achieves faster convergence of the minimum mesh angle and provides better mesh quality after convergence. In this method, bubbles are removed if $verlap > 1.0$; in the original method, bubbles are added or removed if $overlap < 5.0$ or $overlap > 8.0$.

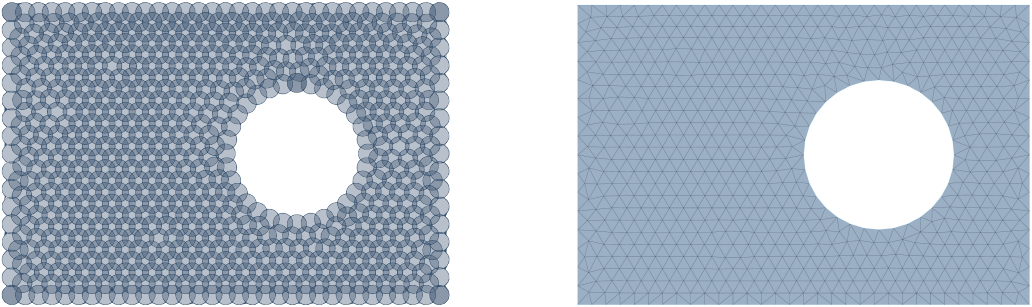

Fig. 8. Uniform bubbles and mesh are generated using the new quantity control method.

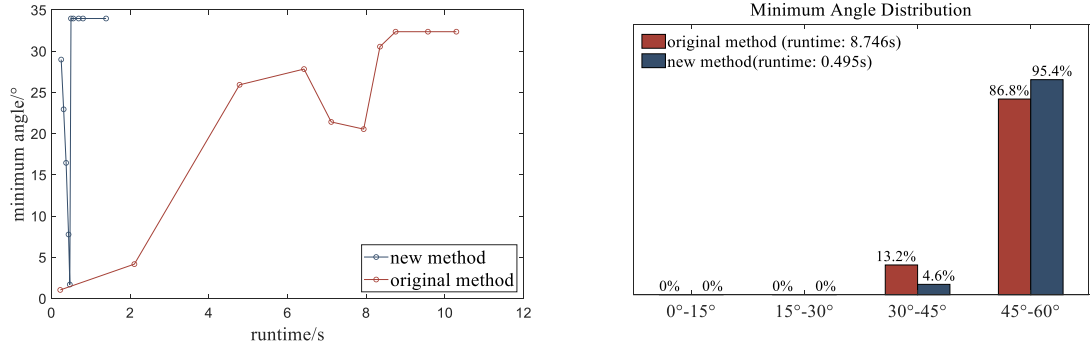

Fig. 9. Comparison of convergence speed and mesh quality.

Fig. 10 shows the non-uniform bubbles and mesh generated on a flat plate using the boundary region quantity control method, with refined mesh near the circular hole. If the overlap thresholds for adding or removing bubbles in the original quantity control method are kept the same as for uniform meshes ($5.0 < overlap < 8.0$), convergence may be excessively slow or even fail. We compared the convergence behavior of the original quantity control method under three different overlap thresholds, as shown in Fig. 11. The results indicate that the convergence speed of the original method is highly sensitive to overlap thresholds, requiring appropriate adjustments according to bubble size variations; otherwise, convergence may be significantly delayed. In contrast, changes to the overlap threshold in the new method have minimal effect on both convergence speed and mesh quality, making it more robust. In this example, the optimal overlap thresholds for the new and original methods are $overlap > 1.2$ and $4.0 < overlap < 10.0$, respectively. The results, shown in Fig. 12, demonstrate that the new quantity control method achieves faster convergence of the minimum mesh angle and produces better mesh quality after convergence.

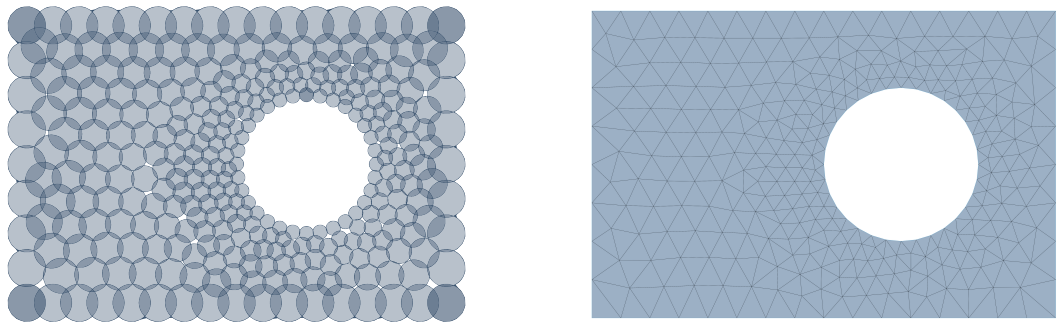

Fig. 10. Uniform bubbles and mesh are generated using the new quantity control method.

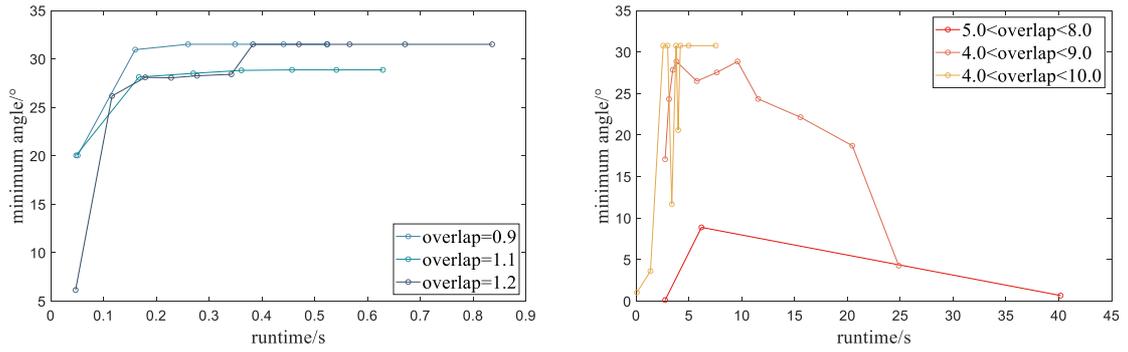

Fig. 11. Effect of overlap threshold on convergence speed.

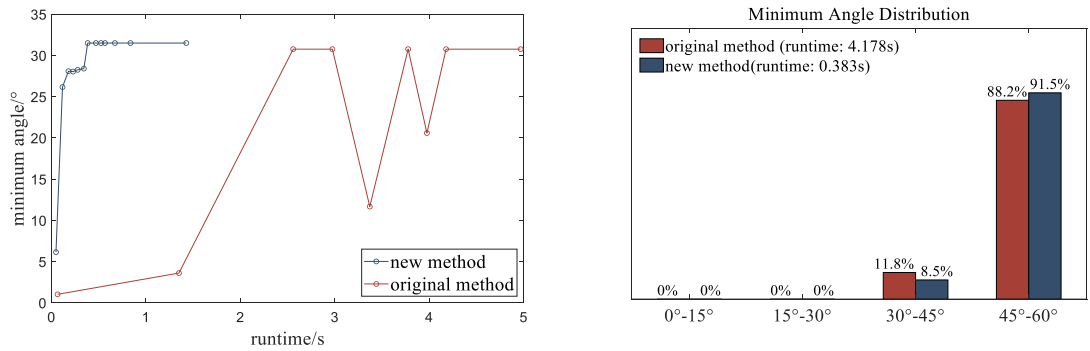

Fig. 12. Comparison of convergence speed and mesh quality.

*Surface Triangulation:*

We performed re-meshing on a curvature-adaptive mesh controlled by the bubble size function in the parametric domain. Fig. 13 shows the initial discrete surface and the re-meshed surface. Table 1 presents the changes in mesh quality and the total time required for surface triangulation. During dynamic bubble adjustment, the minimum angle converged after 18.155 seconds using the new quantity control method, while the original method still had not converged after approximately 100 seconds. We compared the mesh at 104.185 seconds using the original method with the converged mesh from the new method. As shown in Fig. 14, the new method yields better mesh quality.

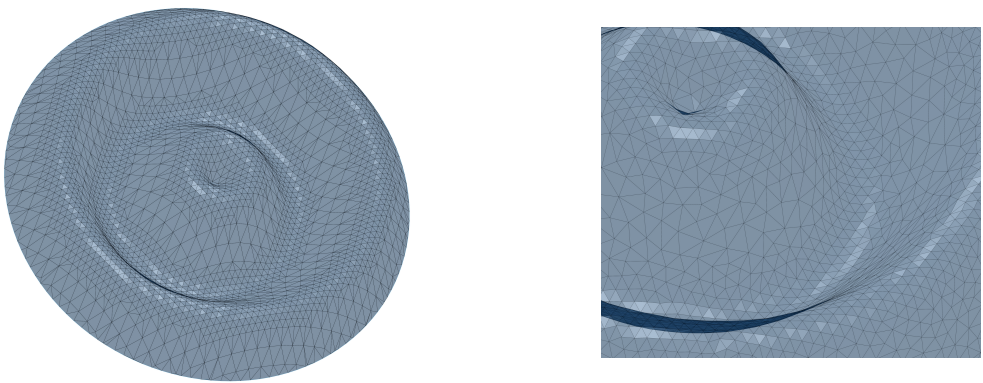

Fig. 13. The initial discrete surface and the re-meshed surface

Table 1. Change in mesh quality after re-meshing

| Run Time | Total Triangles | Minimum Angle | Maximum Angle | Minimum Angle Distribution | | | |
|---|---|---|---|---|---|---|---|
| | | | | 0°~15° | 15°~30° | 30°~45° | 45°~60° |
| 52.269s | 4798 | 11.1322° | 130.017° | 27 | 287 | 702 | 3782 |
| | 6606 | 28.3645° | 111.952° | 0 | 8 | 897 | 5701 |

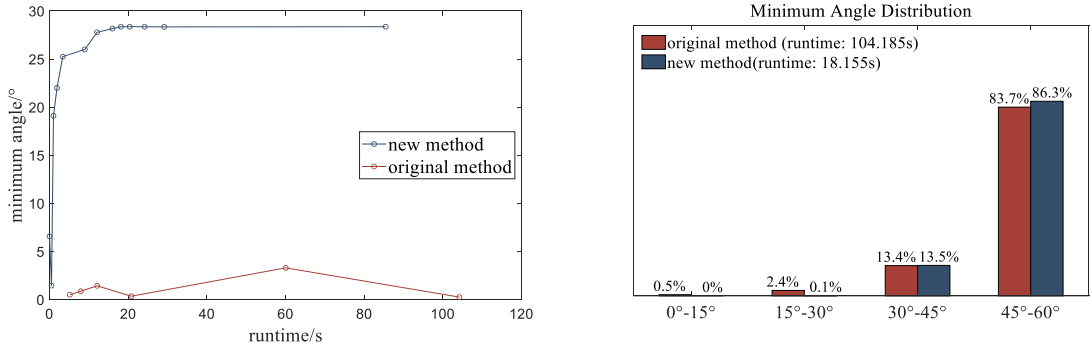

Fig. 14. Comparison of convergence speed and mesh quality.

## 7   Conclusion

The introduction of conformal mapping simplifies the surface bubble packing method by transforming the domain from a surface to a plane, and modifying the bubble quantity control strategy accelerates mesh generation in the planar domain. Numerical experiments show that the improved quantity control method decreases time consumption by over 70%. In surface triangulation, the mesh size can be automatically adjusted to control the error between the discrete surface and the smooth surface within a specified range. In the generated mesh, more than 99% of the triangles have a minimum angle of 30 degrees or greater, and over 80% of the triangles have a minimum angle of 45 degrees or greater.

Further research includes extending the bubble packing method from planar domains to spherical and hyperbolic domains. This extension will enable global triangulation for complex higher genus surfaces without surface cutting, through global conformal parameterization for genus-one and higher genus surfaces.